\documentclass[preprint,12pt]{elsarticle}

\usepackage{amssymb}
\usepackage{amsmath}
\usepackage{comment}

\newcounter{bla}

\journal{Computer Physics Communications}

\begin{document}

\begin{frontmatter}



\title{mqdtfit: A collection of Python functions for empirical multichannel quantum defect calculations}


\author[a]{R. M. Potvliege\corref{cor1}}

\cortext[cor1]{Corresponding author.\\\textit{E-mail address:} r.m.potvliege@durham.ac.uk}
\address[a]{Physics Department, Durham University, South Road, Durham DH1~3LE, UK}

\begin{abstract}
The Python functions distributed with this article can be used for calculating the parameters of multichannel quantum defect theory models describing excited bound states of complex atoms. These parameters are obtained  by fitting a model to experimental data provided by the user. The two main formulations of the theory are supported, namely the one in which the parameters of the model are a set of eigen channel quantum defects and a transformation matrix, and the one where these parameters are the elements of a reactance matrix. The distribution includes programs for calculating theoretical energy levels, calculating mixing coefficients and channel fractions and producing Lu-Fano plots.





\end{abstract}


\end{frontmatter}



\noindent {\bf PROGRAM SUMMARY}

\begin{small}
\noindent
{\em Program Title:} mqdtfit                                         \\
{\em CPC Library link to program files:} (to be added by Technical Editor) \\
{\em Developer's repository link:} https://github.com/durham-qlm/mqdtfit \\
{\em Code Ocean capsule:} (to be added by Technical Editor)\\
{\em Licensing provisions:} BSD 3-clause.  \\
{\em Programming language:} Python~3.                                  \\
{\em Nature of problem:}
Multichannel quantum defect theory aims at giving a unified description of the excited states of multielectron systems whereby the properties of entire series can be predicted from models involving a relatively small number of parameters. These parameters are obtained by fitting theory to experimental data, in the empirical version of the theory. The present programs specifically concern the application of this theory to the bound states of complex atomic systems, such as divalent atoms, for which spectroscopic series are often perturbed by isolated states of a different symmetry, making a multichannel description necessary.\\
{\em Solution method:}
The functions forming this library are grouped and linked to each other into a single Python module. They can be used to calculate the parameters of a given multichannel quantum defect theory model by fitting the model to experimental data provided by the user. These parameters are either a set of eigen channel quantum defects and a transformation matrix, in the eigenchannel parametrization of the theory, or the elements of an orthogonal matrix, in the reactance matrix parametrization of the theory. Given these parameters, this software can also be used to calculate theoretical energy levels, calculate coefficients describing the mixing between the channels considered and produce Lu-Fano plots.
\\
   \\

\end{small}


\section{Introduction}
\label{section:intro}

Multichannel quantum defect theory (MQDT) aims at giving a unified description of the excited states of multielectron systems composed of a single outer electron and a more compact electronic core. This theory makes it possible to understand the properties of entire series of states from models involving a relatively small number of parameters. These parameters are obtained by fitting theory to experiment in the empirical approach we are concerned with, but they may also be calculated from first principles \cite{Aymar1996}.

MQDT has a long history, starting in the early days of Quantum Mechanics \cite{Seaton1983}. This early work has been considerably extended to multichannel cases, initially by Seaton in the 1960s \cite{Seaton1983,Seaton1966}, and later on by a number of other authors. Fano, in particular, pointed out the relevance of the ``eigenchannel" formulation of the theory in a seminal study of the spectrum of molecular hydrogen published in 1970 \cite{Fano1970}. This study and a contemporary analysis of atomic spectra by Lu and Fano \cite{Lu1970} were followed shortly afterwards by further pioneering work, notably by Lu \cite{Lu1971}, and Lee and Lu \cite{Lee1973}.
Apart from more recent extensions, such as the inclusion of various observables other than bound state energies \cite{Aymar1984}, the general approach to obtaining empirical parameters of MQDT models of complex atoms has changed little over the years --- see, e.g., Refs.~\cite{Lehec2018,Robicheaux2018,Robicheaux2019,Zhu2022} for recent examples. Work in that area has often been based on the eigenchannel formulation of MQDT rather on Seaton's original formulation in terms of a ${K}$ (or reactance) matrix. However, the latter is also in use \cite{Robicheaux2018, Robicheaux2019, Cooke1985, Vaillant2014, Vaillantthesis, Vaillant2015,  Potvliege2023}. The calculation yields theoretical energies and mixing coefficients (the $Z_i$ coefficients defined below), which in turn can be used for calculating spontaneous lifetimes and dispersion coefficients --- e.g., \cite{Robicheaux2018, Robicheaux2019, Vaillant2014, Vaillantthesis,Vaillant2015,Aymar1981}.

To the author's knowledge, only one open source, documented, publicly available computer program for empirical MQDT calculations has been published so far, despite the wide relevance of this approach \cite{Robaux1982}. While general in scope, this program does not cover the $K$-matrix formulation of the theory. Moreover, the language in which it is written, Fortran~77, may make it impractical to present-day users. The present publication aims at filling this gap.

Specifically, its purpose is to offer easy-to-use functions for the construction of empirical multichannel quantum defect models of complex atoms. In its present state of development, the code is restricted to models in which all channels are closed, i.e., to bound states. It provides the possibility of fitting the model parameters to a set of experimental data, whether these parameters are expressed in terms of eigen channel quantum defects and a transformation matrix or in terms of a reactance matrix. The module also includes functions calculating energy levels and mixing coefficients and functions plotting channel fractions or Lu-Fano plots, given a set of model parameters. The programming is kept straightforward so as to facilitate customisation and future extensions. Revised versions of the code and new related material will be made available from the URL https://github.com/durham-qlm/mqdtfit (a GitHub repository of the Quantum, Light and Matter research group of Durham University).

The principles of empirical MQDT and the applications of this approach have been reviewed in a number of authoritative articles \cite{Aymar1996,Seaton1983,Aymar1984,Fano1975,Greene2023}, however, without any being sufficiently comprehensive and succinct for serving as a convenient reference for the programs described in this document. These details are reviewed in Section~\ref{section:theory}, for completeness and clarity. Practical advice about fitting MQDT models to experimental spectra can be found, e.g., in Refs.~\cite{Lu1971,Lee1973,Robaux1982,Armstrong1977}.
The numerical methods used in this implementation of MQDT theory and other specific details are outlined in Section~\ref{section:methods}. Further information about the use of this module can be found in Section~\ref{section:generalinfo}. A more detailed user manual is joined with the distribution, as are example programs and their output. The latter are briefly described in Section~\ref{section:examples}. 

\section{Multichannel quantum defect theory of bound systems}
\label{section:theory}

Multichannel quantum defect theory concerns systems in which several different configurations of the core need to be considered, hence different configurations of the (core - outer electron) system. Each of these configurations is referred to as a channel. We are primarily interested by the dissociation channels, which are characterised by the angular momentum of the outer electron and by the state the core would be in if this electron was infinitely far away from it (dissociation channels are also called scattering channels). We will assume that we need to consider $N_{\rm ch}$ such channels and denote by $\mathfrak{I}_i$ the ionisation limit of channel $i$ ($i = 1,\ldots,N_{\rm ch}$): in a state of energy $\mathfrak{E}$, channel $i$ is open or closed according to whether $\mathfrak{E} > \mathfrak{I}_i$ or $\mathfrak{E} < \mathfrak{I}_i$. We reserve the symbols $E$ and $I_i$ for the corresponding energies expressed as wave numbers:
\begin{equation}
    E \equiv \mathfrak{E}/hc, \qquad I_i \equiv \mathfrak{I}_i/hc,
\end{equation}
where $h$ is Planck's constant.

The programs described here specifically address the case where all dissociation channels are closed. The theory is based on two ans{\"a}tze, namely
\begin{equation}
\Psi(X,r)= {\cal A}\sum_{i}^{} C_i \chi_i(X) \phi_i(r)
\label{eq:totalwavefunction1}
\end{equation}
and 
\begin{equation}
    \Psi(X,r) = {\cal A}\sum_\alpha A_\alpha \left[{\cal  F}_\alpha(X,r) \cos(\pi\mu_\alpha) - {\cal G}_\alpha(X,r) 
    \sin(\pi\mu_\alpha)\right],
    \label{eq:Psirewritten}
\end{equation}
where $r$ denotes the radial coordinate of the outer electron, $X$ represents all the other relevant coordinates and ${\cal A}$ is the antisymmetrization operator.
These two equations both express the wave function of an energy eigenstate of interest as an antisymmetrized linear combination of product wave functions; however, they do so in different ways: Eq.~(\ref{eq:totalwavefunction1}) expresses the wave function directly in terms of the individual dissociation channels, whereas Eq.~(\ref{eq:Psirewritten}) expresses it in terms of configurations better suited to describing the inner region where all the electrons are close to each other and to the nucleus. This other form of the wave function appears naturally in calculations based on the diagonalisation of the (core + outer electron) Hamiltonian in this region. The $\alpha$-configurations are usually referred to as the eigenchannels and the constants $\mu_\alpha$ as the corresponding eigen quantum defects.

In Eq.~(\ref{eq:totalwavefunction1}), the $C_i$'s are constant coefficients describing the admixture of the different dissociation channels in the state $\Psi$, and, as implied by the notation, the $\phi_i$'s are functions depending only on the radial coordinate of the outer electron
while the $\chi_i$'s are functions of the angular coordinates of the outer electron, of the coordinates of the other electrons and of the spin of all the electrons.
Making the approximation that the outer electron moves in a Coulomb field when outside the core leads to the following equation, in the case of a neutral atom:
\begin{equation}
  \left[ -\frac{\hbar^2}{2\mu}\left(\frac{d^2}{dr^2} - \frac{l_i(l_i+1)}{r^2}\right) - \frac{e^2}{4\pi\epsilon_0 r} \right] u_i(r) = \epsilon_i u_i(r),
\label{eq:Coulombequation}
\end{equation}
where $u_i(r) = r\phi_i(r)$, $\mu$ is the reduced mass, $e$ is the charge of the electron and $\epsilon_i$ is an energy measured relative to $\mathfrak{I}_i$. For a state of energy $\mathfrak{E}$,
\begin{equation}
    \epsilon_i = \mathfrak{E} - \mathfrak{I}_i = (E - I_i)\,hc.
    \label{eq:epsilonidef}
\end{equation}
The general solution of Eq.~(\ref{eq:Coulombequation}) can be expressed in terms of a linear combination of regular and irregular Coulomb functions,\footnote{We take these functions to be the $f$ and $g$ functions defined, e.g., in Refs.~\cite{Fano1970}, \cite{Greene1979} and \cite{Gallagher1994}, not the $f$ and $g$ functions defined by Seaton~\cite{Seaton1983,Seaton1966}. In the definition adopted here, these functions are identical to the functions denoted $\mathfrak{f}$ and $\mathfrak{g}$ by Lee and Lu~\cite{Lee1973} and by Dehmer and Fano~\cite{Dehmer1970}. They are almost identical to Seaton's functions $s$ and $c$~\cite{Seaton1983,Seaton1966}, which are normalised differently: $f \equiv \sqrt{2} s$ and $g \equiv -\sqrt{2} c$.}
$f(\epsilon,l;r)$ and $g(\epsilon,l;r)$. We can thus write
\begin{equation}
u_i(r) = f(\epsilon_i,l_i;r) \cos \pi\nu_i + g (\epsilon_i,l_i;r) \sin\pi\nu_i,
\label{eq:longrange}
\end{equation}
where $\nu_i$ is a constant.

In Eq.~(\ref{eq:Psirewritten}), on the other hand, the $A_\alpha$'s are constant coefficients whereas the ${\cal F}_\alpha$'s and ${\cal G}_\alpha$'s are functions of the coordinates and spin of all the electrons: the ${\cal F}_\alpha$'s are assumed to reduce to linear combinations of the functions $\chi_i(X) \,f(\epsilon_i,l_i;r)/r$ and the 
${\cal G}_\alpha$'s to linear combinations of the functions $\chi_i(X) \,g(\epsilon_i,l_i;r)/r$ when the outer electron is far from from the core. Specifically, we assume that, for sufficiently large values of $r$,
\begin{align}
    {\cal F}_\alpha(X,r) &= \sum_i U_{i\alpha}\,\chi_i(X) \,f(\epsilon_i,l_i;r)/r,
    \label{eq:Falpha}\\
    {\cal G}_\alpha(X,r) &= \sum_i U_{i\alpha}\,\chi_i(X), g(\epsilon_i,l_i;r)/r,
    \label{eq:Galpha}
\end{align}
where the coefficients $U_{i\alpha}$ are real and do not depend on electron coordinates.

The values of $\nu_i$, $C_i$, $\mu_\alpha$ and $A_\alpha$ for which these two ans{\"a}tze yield physically admissible bound state wave functions are restricted by three requirements, namely
(1) that $\phi_i(r)$ goes to 0 for $r \rightarrow \infty$;
(2) that Eqs.~(\ref{eq:totalwavefunction1}) and (\ref{eq:Psirewritten}) apply not only to the bound states of interest but also to the states with energy above the lowest ionisation threshold, for which some or all the dissociation channels are open;
and (3) that
\begin{equation}
\sum_{i}^{} C_i \chi_i(X) \phi_i(r)
 \equiv \sum_\alpha A_\alpha \left[{\cal  F}_\alpha(X,r) \cos(\pi\mu_\alpha) - {\cal G}_\alpha(X,r) 
    \sin(\pi\mu_\alpha)\right]
    \label{eq:condition}
\end{equation}
when $r$ is sufficiently large.

The first of these requirements implies that the $\nu_i$'s are necessarily related to the $\epsilon_i$'s by the equation \cite{Seaton1983}
\begin{equation}
\epsilon_i = -\frac{\mu e^4}{(4\pi\epsilon_0)^2\hbar^2}\,\frac{1}{2\nu_i^2},
\quad i=1,\ldots,{N}_{\rm ch}.
\label{eq:energies}
\end{equation}
(By analogy with the case of atomic hydrogen, the $\nu_i$'s are often referred to as effective principal quantum numbers.) In terms of energies expressed as wave numbers,
\begin{equation}
    E - I_i = -\tilde{R}/\nu_i^2,
    \quad i=1,\ldots,{N}_{\rm ch},
    \label{eq:nudefined}
\end{equation}
where $\tilde{R}$ is the mass-corrected Rydberg constant for the species considered.

The second requirement implies, as an indirect consequence of the unitarity of the scattering matrix, that the square matrix $[U_{i\alpha}]$ of elements $U_{i\alpha}$ is orthogonal --- see, e.g., Refs.~\cite{Seaton1983}, \cite{Lu1971} or \cite{Greene2023}.

The last requirement has several consequences, as is explained in \ref{section:appendix}. One finds, in particular, that the coefficients $C_i$ and $A_\alpha$ satisfy the following systems of equations:
\begin{alignat}{3}
&\sum_{i}^{} U_{i \alpha} \sin[ \pi (\nu_i + \mu_\alpha) ]C_i &&=0, \quad \alpha &&=1,\ldots,{N}_{\rm ch},
\label{eq:Cisystem}\\
&\sum_{\alpha}^{} U_{i \alpha} \sin[ \pi (\nu_i + \mu_\alpha) ]A_\alpha &&=0, \quad i &&=1,\ldots,{N}_{\rm ch}.
\label{eq:Aalphasystem}
\end{alignat}
Moreover,
\begin{alignat}{3}
A_\alpha &= \sum_{i}^{} U_{i\alpha}\cos[\pi(\nu_i + \mu_\alpha)]C_i, \quad  &&\alpha &&= 1,\ldots,N_{\rm ch}
\label{eq:AfromC}
\end{alignat}
and
\begin{alignat}{3}
C_i &= \sum_{\alpha}^{} U_{i\alpha}\cos[\pi(\nu_i + \mu_\alpha)]A_\alpha, \quad  &&i &&= 1,\ldots,N_{\rm ch}.
\label{eq:CfromA}
\end{alignat}
It is necessary, for Eqs.~(\ref{eq:Cisystem}) and (\ref{eq:Aalphasystem}) to have non-trivial solutions, that
\begin{equation}
\det | U_{i \alpha} \sin[\pi (\nu_i + \mu_\alpha)] | =0.
\label{eq:eigendetcondition}
\end{equation}
Given a $[U_{i\alpha}]$ matrix and a set of eigen quantum defects $\mu_\alpha$, this last equation determines the $\nu_i$'s, thus the bound state energies. 

Alternatively (see \ref{section:appendix}), the coefficients $C_i$ can also be obtained by solving a related system of linear equations, namely
\begin{equation}
\sum_{j}^{} \left[ K_{ij} + \delta_{ij} \tan(\pi \nu_i) \right] c_j = 0, \quad i=1,\ldots,{N}_{\rm ch},
\label{eq:Kmatrixsystem}
\end{equation}
where the $K_{ij}$ coefficients are defined by the matrix equation
\begin{equation}
[K_{ij}] = [U_{i\alpha}] [\delta_{\alpha \alpha'} \tan(\pi \mu_\alpha)] [U_{j \alpha'}]^\dagger
\label{eq:eigentoK}
\end{equation}
and the $c_j$'s are related to the $C_j$'s by the equation
\begin{equation}
C_j = c_j / \cos(\pi \nu_j).
\label{eq:citoCi}
\end{equation}
The system (\ref{eq:Kmatrixsystem}) admits a non-trivial solution under the condition that
\begin{equation}
\det | K_{ij} + \delta_{ij}\tan(\pi \nu_i) | = 0.
\label{eq:Kmatrixdetcondition}
\end{equation}
This last equation replaces the condition (\ref{eq:eigendetcondition}) determining the possible values of the $\nu_i$'s, in this alternative formulation. The bound states energies are then parametrized by the matrix $[K_{ij}]$ rather than by the matrix $[U_{i\alpha}]$ and the eigen quantum defects $\mu_\alpha$. The two formulations are equivalent as long as Eq.~(\ref{eq:eigentoK}) holds.
By analogy with similar equations appearing in scattering theory, the matrix $[K_{ij}]$ is often called the reactance (or reaction) matrix. As can be seen from Eq.~(\ref{eq:eigentoK}), this matrix is symmetric ($K_{ij} = K_{ji})$, its eigenvalues are $\tan(\pi\mu_\alpha)$, $\alpha = 1,\ldots,N_{\rm ch}$, and its eigenvectors are the corresponding columns of the $[U_{i\alpha}]$ matrix.

The ionization thresholds $I_i$, the $[K_{ij}]$ matrix, the $[U_{i\alpha}]$ matrix and the eigen quantum defects can, in principle, be calculated from first principles. However, in the empirical approach we follow here, the $I_i$'s are derived directly from measurements and the other parameters are obtained by fitting the theoretical energies to spectroscopic data, assuming that these quantities vary slowly with $E$. Either the $K_{ij}$'s or the $U_{i\alpha}$'s and $\mu_\alpha$'s are directly obtained by fitting, giving two alternative formulations of empirical MQDT.

It can be noted that the functions $u_i(r)$ of Eq.~(\ref{eq:longrange}) are proportional to exponentially decreasing Whittaker functions when Eq.~(\ref{eq:energies}) is fulfilled. Namely, for such values of $\nu_i$ \cite{Seaton1983},
\begin{equation}
u_i(r) = (-1)^{l_i+1}\nu_i^{3/2}\,D(\nu_i,l_i)\,W_{\nu_i,l_i+1/2}(2r/\nu_i)
\end{equation}
with
$D(\nu_i,l_i) = [\nu_i^2\Gamma(\nu_i+l_i+1)\Gamma(\nu_i-l_i)]^{-1/2}$.
Following Seaton \cite{Seaton1983}, Eq.~(\ref{eq:totalwavefunction1}) is usually recast in terms of these functions, as
\begin{equation}
\Psi= {\cal A}\sum_{i}^{} Z_i \chi_i \,D(\nu_i,l_i)\,W_{\nu_i,l_i+1/2}(2r/\nu_i)/r, 
\label{eq:totalwavefunction2}
\end{equation}
where
\begin{align}
Z_i &= (-1)^{l_i+1} \nu_i^{3/2}\,C_i  
= (-1)^{l_i+1} \nu_i^{3/2}\,\sum_\alpha U_{i\alpha}\cos[\pi(\nu_i + \mu_\alpha)]A_\alpha.
\end{align}
The $C_i$'s and $A_\alpha$'s, and thus the $Z_i$'s, are defined only to within an overall constant factor by Eqs.~(\ref{eq:Cisystem}),  (\ref{eq:Aalphasystem}) and (\ref{eq:Kmatrixsystem}). The value of this factor is constrained by the normalisation of the wave function, $\Psi(X,r)$: its value must be such that the $Z_i$'s satisfy the condition \cite{Seaton1983}
\begin{equation}
    \sum_{i,j}^{} Z_i M_{ij} Z_j = 1,
\label{eq:normalisation0}
\end{equation}
where\footnote{Note that the energy ${\cal E}$ appearing in Eq.~(6.50) of Ref.~\cite{Seaton1983} is the ``Z-scaled energy", ${\cal E} = 2\mathfrak{E}/Z^2$ where $Z = 1$ for a neutral atom.
}
\begin{align}
M_{ij} &= \delta_{ij} + \frac{1}{2}\,
\frac{\mu e^4}{(4\pi\epsilon_0)^2\hbar^2}\,
q_i \left(\frac{{\rm d}K_{ij}}{{\rm d}\mathfrak{E}}\right) q_j\nonumber\\
&= \delta_{ij} + \tilde{R}\,
q_i \left(\frac{{\rm d}K_{ij}}{{\rm d}{E}}\right) q_j
\label{eq:normalisation}
\end{align}
with
\begin{equation}
    q_i = (-1)^{l_i} \left(\frac{2}{\pi\nu_i^3}\right)^{1/2} \cos(\pi\nu_i).
\end{equation}
The mixing coefficients $Z_i$ and the channel fractions $|Z_i|^2$ quantify the importance of each dissociation channel in the states of interest.
The energy dependence of ${K}$ is often neglected in the normalisation condition, in which case the channel fractions are taken to sum to unity:
\begin{equation}
    \sum_{i}^{} |Z_i|^2 = 1.
    \label{eq:normalisationsimplified}
\end{equation}

It is often preferable to use different angular momentum coupling schemes for the dissociation channels than for the eigenchannels, as the latter refer to a region where the outer electron interacts more strongly with the core than in the former. Transforming the mixing coefficients from the coupling scheme used for the dissociation channels to one better adapted to the eigenchannels can be done by way of an orthogonal transformation matrix, $[U_{i\bar{\alpha}}]$, where the index $\bar{\alpha}$ ($\bar{\alpha} = 1,\ldots,N_{\rm ch}$) labels the channels in the transformed coupling scheme:
\begin{equation}
    Z_{\bar{\alpha}} = \sum_i U_{i\bar{\alpha}}Z_i.
\end{equation}
In the case of complex atoms, for example, LS-coupling is often better suited for describing the eigenchannels, since all the electrons interact strongly in the inner region, whereas jj-coupling is often better suited for describing the dissociation channels, since the outer electron is nearly free when far from the core. The recoupling matrix can be written in terms of Wigner 9-j symbols in this case \cite{Lee1973}:
\begin{align}
U_{i\bar{\alpha}}&= \sqrt{(2j_{{c}i}+1)(2j_{{e}i}+1)(2L_{\bar{\alpha}}+1)(2S_{\bar{\alpha}}+1)}\;\begin{Bmatrix}
l_{ci} & s_{ci} & j_{ci}\\
l_{ei} & s_{ei} & j_{ei}\\
L_{\bar{\alpha}} & S_{\bar{\alpha}} & J_{\bar{\alpha}}
\end{Bmatrix},
\label{eq:recoupling}
\end{align}
where $l_{ci}$, $s_{ci}$ and $j_{ci}$ are the quantum numbers of the core in channel $i$, $l_{ei}$, $s_{ei}$ and $j_{ei}$ the quantum numbers of the outer electron in channel $i$, and $L_{\bar{\alpha}}$, $S_{\bar{\alpha}}$ and $J_{\bar{\alpha}}$ the quantum numbers of the LS-coupled system.

The matrix $[U_{i\alpha}]$ is usually written in the form of a product of the transformation matrix $[U_{i\bar{\alpha}}]$ and another orthogonal matrix, $[V_{{\bar \alpha}\alpha}]$, the latter being itself factorised into a product of $N_{\rm rot}$
rotation matrices $[R_{\bar{\alpha}\alpha}(m)]$:
\begin{equation}
[U_{i\alpha}] = [U_{i\bar{\alpha}}][V_{\bar{\alpha}\alpha}]
 = [U_{i\bar{\alpha}}][R_{\bar{\alpha}\alpha}(1)]\,
[R_{\bar{\alpha}\alpha}(2)] \cdots [R_{\bar{\alpha}\alpha}({N}_{\rm rot})].
\label{eq:UandV}
\end{equation} 
The rotation matrices $[R_{\bar{\alpha}\alpha}(m)]$ are defined as follows: each $m$ corresponds to an angle $\theta_m$ and to a pair of indices, $p_m$ and $q_m$, and
\begin{equation}
  R_{\bar{\alpha}\alpha}(m) = \begin{cases}
  \cos \theta_m & \mbox{if}\;\bar{\alpha} = \alpha = p_m,\\
 -\sin \theta_m & \mbox{if}\;\bar{\alpha} = p_m\;\mbox{and}\; \alpha = q_m,\\
   \sin \theta_m & \mbox{if}\;\bar{\alpha} = q_m\;\mbox{and}\; \alpha = p_m,\\
   \cos \theta_m & \mbox{if}\;\bar{\alpha} = \alpha = q_m,\\
    1 & \mbox{if}\;\bar{\alpha} = \alpha\not= p_m, q_m, \\ 
   0& \mbox{for any other values of}\;\bar{\alpha}\;\mbox{and}\;\alpha.
  \end{cases}
\end{equation}
Since $[V_{{\bar \alpha}\alpha}]$ is an $N_{\rm ch}\times N_{\rm ch}$ orthogonal matrix, at most ${N}_{\rm ch}({N}_{\rm ch}-1)/2$ rotations are necessary for factorizing $[V_{{\bar \alpha}\alpha}]$ in that way.
The angles $\theta_m$ are taken to be fitting parameters in empirical MQDT calculations carried out along these lines, together with the eigen quantum defects $\mu_\alpha$. In practice, fewer than ${N}_{\rm ch}({N}_{\rm ch}-1)/2$ are usually necessary for achieving a suitable fit. The factorisation therefore has the double advantage of making it possible to reduce the number of fitting parameters and of ensuring that the resulting $[U_{i\alpha}]$ matrix is orthogonal. 

\section{Methods}
\label{section:methods}
\subsection{Energy dependence of the MQDT parameters}

The elements of the $[K_{ij}]$ matrix as well as those of the $[V_{\bar{\alpha}\alpha}]$ matrix and the eigen quantum defects $\mu_\alpha$ are, in principle, energy-dependent. However, it is found that this dependence is usually slow and does not need to be described to a high level of accuracy in order to obtain bound states energies in satisfactory agreement with experiment.

Standard practice in calculations based on the eigenchannel version of the theory is to take
$[V_{\bar{\alpha}\alpha}]$ to be constant in energy and assume a linear dependence on $E$ for the eigen quantum defects.
Following previous work in that area --- see, e.g., Refs.~\cite{Robaux1982} and \cite{Esherick1977} --- two alternative parametrisations are implemented in the present programs: each energy-dependent $\mu_\alpha$ is written in terms of two constants, $\mu_\alpha^{(0)}$ and $\mu_\alpha^{(1)}$, and either
\begin{equation}
    \mu_\alpha = \mu_\alpha^{(0)} + \mu_\alpha^{(1)}\, \frac{(I_s - E)}{I_s},
    \label{eq:energydependencemua}
\end{equation}
where $I_s$ is the first ionisation threshold, or
\begin{equation}
    \mu_\alpha = \mu_\alpha^{(0)} + \mu_\alpha^{(1)}\, \frac{(I_s - E)}{\tilde R}.
    \label{eq:energydependencemub}
\end{equation}
Neglecting the dependence on $E$ of the matrix $[V_{\bar{\alpha}\alpha}]$ makes it possible to reduce Eqs.~(\ref{eq:normalisation0}) and (\ref{eq:normalisation}) to the simpler form~\cite{Lee1973}
\begin{equation}
\sum_{i} Z_i^2 + 
2\tilde{R}
\,\sum_\alpha
\left(\frac{{\rm d}\mu_\alpha}{{\rm d}E}\right) A_\alpha^2 = 1.
\label{eq:normalisationeigen}
\end{equation} 

The calculations based on the ${K}$-matrix version of the theory follow Ref.~\cite{Vaillant2014}, in that this matrix is allowed to vary with energy through a linear dependence of its diagonal elements, its off-diagonal elements being taken to be constant. As for the eigenchannel formulation, two alternative parametrisations of this dependence are implemented in this module, one in which the diagonal
elements are written as
\begin{equation}
K_{i i} = K_{i i}^{(0)} + K_{i i}^{(1)}  \frac{(I_s - E)}{I_s},
\label{eq:energydependenceKa}
\end{equation}
and one in which they are written as
\begin{equation}
K_{i i} = K_{i i}^{(0)} + K_{i i}^{(1)}  \frac{(I_s - E)}{\tilde R},
\label{eq:energydependenceKb}
\end{equation}
the coefficients $ K_{i j}^{(0)}$ and $K_{i j}^{(1)}$ being constant. The off-diagonal elements of the ${K}$-matrix are assumed to be constant:
\begin{equation}
    K_{ij} \equiv K_{ij}^{(0)},\quad i \not=j.
\end{equation}

\subsection{Calculation of theoretical energies}
\label{section:energies}

The energies are determined by the condition that Eqs.~(\ref{eq:nudefined}) and either Eq.~(\ref{eq:eigendetcondition}) or Eq.~(\ref{eq:Kmatrixdetcondition}) are satisfied simultaneously, for a given ${K}$-matrix or for given eigen quantum defects and rotation angles. One of the channels, channel~$j$, is given a special role in the calculation. Typically, channel $j$ would be chosen amongst those dominating the series of interest; however, this choice is largely immaterial. In view of Eqs.~(\ref{eq:nudefined}), a given value of $\nu_j$ corresponds to a value of $\nu_i$ for the $i$-th channel given by the equation
\begin{equation}
\nu_i = F_i(j;\nu_j),
\label{eq:nujtonui}
\end{equation}
where
\begin{equation}
F_i(j;\nu_j)= \left[ \frac{I_i - I_j}{\tilde{R}} + \frac{1}{\nu_j^{2}} \right]^{-{1}/{2}}.
\label{eq{Fdefined}}
\end{equation}
The bound state energies are found by identifying the values of $\nu_j$ for which Eq.~(\ref{eq:eigendetcondition}) or Eq.~(\ref{eq:Kmatrixdetcondition}) is satisfied when all the other $\nu_i$'s are calculated as per Eq.~(\ref{eq:nujtonui}). The program offers the possibility of finding these values through a direct search of the roots of the relevant determinant or through a more indirect method outlined below.

For efficiency, the direct search is confined to the vicinity of values of $\nu_j$ corresponding to experimental energies: given a set of experimental energies $E_n^{\,\rm exp}$, the program searches for a root of the determinant in each of the intervals 
\begin{displaymath}
\nu_{j,n}^{\rm exp} - \delta \nu_n^{(-)} \leq \nu_{j} \leq \nu_{j,n}^{\rm exp} + \delta \nu_n^{(+)},
\end{displaymath}
where
\begin{equation}
\nu_{j,n}^{\rm exp} = [(I_j-E_n^{\,{\rm exp}})/\tilde{R}]^{-1/2}.
\label{eq:nujexp}
\end{equation}
Two different schemes are possible for this: the $\delta \nu_n^{(+)}$'s and 
$\delta \nu_n^{(-)}$'s can be chosen to be multiples of a single user-defined $\delta\nu$ or to vary from energy to energy.

For each $E_n$, in the first scheme, the program explores a sequence of search intervals $\Delta_n(0)$, $\Delta_n(-1)$, $\Delta_n(1)$, $\Delta_n(-2)$, $\Delta_n(2)$, etc.,
until one is found to contain a root of the determinant. Here
\begin{equation}
\Delta_n(p) = [\nu_{j,n}^{\rm exp} - \delta\nu + p\,\delta\nu\,,\, \nu_{j,n}^{\rm exp} + \delta\nu + p\,\delta\nu], \quad p = 0,\pm 1, \pm 2,\ldots
\label{eq:aoption}
\end{equation}
$\Delta_n(p)$ is deemed not to bracket a root if the sign of the determinant is the same at $\nu_{j,n}^{\rm exp} - \delta\nu + p\,\delta\nu$ as at $\nu_{j,n}^{\rm exp} + \delta\nu + p\,\delta\nu$. Once a bracketing interval is found, a root is narrowed down by the Brent's method based {\tt brentq} function of the {\tt scipy.optimize} library.

The other scheme determine the $\delta \nu_n^{(+)}$'s and 
$\delta \nu_n^{(-)}$'s from the distance between the different experimental energies. Assuming that these energies are ordered by increasing values, $\delta \nu_n^{(+)}$ and 
$\delta \nu_n^{(-)}$ are taken such that
\begin{equation}
    \delta \nu_n^{(+)} = \alpha \left(\nu_{j,n+1}^{\rm exp} - \nu_{j,n}^{\rm exp}\right), \qquad
     \delta \nu_n^{(-)} = \alpha \left(\nu_{j,n}^{\rm exp} - \nu_{j,n-1}^{\rm exp}\right),
\label{eq:roption}
\end{equation}
where $\alpha$ is a user-defined numerical factor. These intervals do not vary in the course of the calculation, and, as above, the {\tt brentq} function is used to locate the roots of the determinant.

As an alternative on a direct search for the roots of the relevant determinant, the program also implements the method described in Ref.~\cite{Vaillant2014}, in which the problem is reformulated in terms of a search for the minima of a function. Two channels play a special role in this method, channel $j$ as above, and channel $k$, which must have a different ionisation limit than channel $j$ but whose choice is otherwise arbitrary. Given a value of $\nu_j$ and the corresponding values of $\nu_i$ for the channels for which $I_i \not= I_k$, those $\nu_i$'s being  obtained from Eq.~(\ref{eq:nujtonui}), the left-hand side of Eq.~(\ref{eq:Kmatrixdetcondition}) is a polynomial of degree $N_k$ in $\tan(\pi\nu_k)$, where $N_k$ is the number of dissociation channels converging to the same ionisation limit as channel $k$ ($N_k = 1$ if channel $k$ is the only channel with that ionisation limit). Calculating the roots of this polynomials yields $N_k$ different $\nu_j$-dependent values of $\nu_k$ in the interval $[0,1)$, i.e., $G_{k;1}(\nu_j)$, $G_{k;2}(\nu_j)$, \ldots , $G_{k;N_k}(\nu_j)$.\footnote{Due to the periodicity of the tangent function, the system is also satisfied for $\nu_k = G_{k;s}(\nu_j) \pm 1$, $ G_{k;s}(\nu_j)\pm 2$, etc. However, these other solutions are not relevant for our purposes.}
One of these values must coincide with $F_k(\nu_k)\;\mbox{mod}\, 1$ at those values of $\nu_j$ corresponding to a bound state energy. Solving Eq.~(\ref{eq:Kmatrixdetcondition}) thus amounts to searching for the zeros of the functions $G_{k;s}(\nu_j) - F_k(\nu_j)\;\mbox{mod}\, 1$, $s=1,\ldots,N_k$.
It has been noticed \cite{Vaillantthesis} that this problem is best tackled by searching for the minima of the function $\Xi^2(\nu_j)$, where
\begin{equation}
    \Xi^2(\nu_j) \equiv \min \,[G_{k;s}(\nu_j) - F_k(\nu_j)\;\mbox{mod}\, 1]^2,\quad s=1,\ldots,N_k.
\end{equation}

The calculation is programmed accordingly, using the implementation of the Brent's minimisation method offered by the function {\tt minimize\_scalar} of the {\tt scipy.optimize} library. The search intervals are as described above, with the difference that their bounds are meant to bracket a minimum of $\Xi^2(\nu_j)$ rather than bracket a root of the relevant determinant. As mentioned in the previous paragraph, calculating the $G_{k;s}(\nu_j)$'s involves representing $\det | K_{ij} + \delta_{ij}\tan(\pi \nu_i) |$ as a polynomial of order $N_k$ in $\tan(\pi\nu_k)$. The $(N_k + 1)$
coefficients of this polynomial are obtained by calculating the determinant at $(N_k + 1)$ different values of $\nu_k$ \cite{Robaux1982} and using the {\tt polyfit} function of the {\tt polynomial.polynomial} module of the {\tt numpy} library to fit a polynomial of order $N_k$ to these values. The function {\tt polyroots} of this module is then used to calculate the roots of that polynomial. The required elements of the ${K}$-matrix are obtained from the eigen quantum defects and the rotation angles, as indicated by Eqs.~(\ref{eq:eigentoK}) and (\ref{eq:UandV}), in calculations based on the eigenchannel formalism.

\subsection{Optimization of the MQDT parameters}
\label{section:optimization}

The MQDT parameters ({$K$}-matrix elements, or eigen quantum defects and rotation angles) can be optimized by $\chi^2$-fitting to a set of experimental energies. These parameters are divided into fitting parameters, whose values are variable, and static parameters, whose values (normally 0) remain constant. Given a set of experimental energies $E_n^{\,\rm exp}$ and associated experimental errors $\alpha_n^{\rm exp}$, the program calculates the corresponding theoretical energies $E_n^{\,\rm th}$ as described in Section~\ref{section:energies}
and varies the fitting parameters so as to minimise the value of $\chi^2$, where
\begin{equation}
\chi^2= \sum_n [(E_n^{\,\rm exp} - E_n^{\,\rm th})/\alpha_n^{\rm exp}]^2.
\label{eq:chisquared}
\end{equation}
A Nelder-Mead downhill simplex algorithm is used to this effect, through the function {\tt fmin} of the {\tt scipy.optimize} library. The user must provide starting values of the relevant MQDT parameters.

In calculations based on the eigenchannel formulation of the theory, the $\mu_{\alpha}^{(0)}$'s and the $\theta_m$'s are always assumed to be variable fitting parameters, whereas each of the $\mu_\alpha^{(1)}$'s can be taken to be a static parameter (set to a given invariable values) or a variable fitting parameter, to the choice of the user. In calculations based on the ${K}$-matrix formulation of the theory, each of the $K_{ij}^{(0)}$'s and of the $K_{ii}^{(1)}$'s can be defined as being static or variable.

\subsection{Lu-Fano plots}
\label{section:Lu_Fano}

The program includes a function producing Lu-Fano plots \cite{Lu1970}, i.e., here, plots of $\nu_k \;\mbox{mod}\, 1$ vs.\ $\nu_j$, where $\nu_j$ and $\nu_k$ are the same effective principal quantum numbers as in Section~\ref{section:energies}. The experimental data are indicated by markers located at the points of co-ordinates $(\nu_{j,n}^{\rm exp},\nu_{k,n}^{\rm exp})$, where
the $\nu_{j,n}^{\rm exp}$'s are defined by Eq.~(\ref{eq:nujexp}) and the $\nu_{k,n}^{\rm exp}$'s by the equation
\begin{equation}
\nu_{k,n}^{\rm exp} = [(I_k-E_n^{\,{\rm exp}})/\tilde{R}]^{-1/2}.
\label{eq:nukexp}
\end{equation}
Theory curves are also plotted, showing how $\nu_k$ varies as a function of $\nu_j$ when $\nu_k$ is taken equal to $G_{k;s}(\nu_j)$, $s=1,\ldots,N_k$. These $N_k$ curves pass through the data points when the MQDT model describes the experimental energies well.

\section{General information about the module}
\label{section:generalinfo}

\subsection{Installation and general comments}

The programs forming this library implement the theory and methods outlined in Section~\ref{section:theory} of this document in a generally straightforward way. These programs are entirely written in Python~3 and are grouped in a single file, forming a single module. This module can be made available to other programs by an {\tt import} statement. Specifically, the functions provided by this module can be used (1) to calculate theoretical energies, mixing coefficients and channel fractions given the details of an MQDT model and a set of experimental energies; (2) to optimize the parameters of an MQDT model; (3) to evaluate the $\chi^2$ statistics characterizing the goodness of fit of an MQDT model to a set of experimental energies; and (4) to draw Lu-Fano plots.

The module currently contains 40 functions, amongst which 14 are intended to interface with a driving program provided by the user. A technical description of the input parameters of each of these functions and of the results they return can be found at the start of the respective code. A brief description of each of the user-facing functions is given in a user manual provided with the distribution. Examples of driving programs are also provided and are briefly described in Section~\ref{section:examples}.

\subsection{Using the module}

Any use of the module must start by a call to either one of two initialisation functions, {\tt initialize\_eigenchannel} or {\tt initialize\_Kmatrix}, respectively for calculations in the eigenchannel formulation of the theory and for calculations formulated in terms of a ${K}$-matrix. Calculations in one formulation cannot be mixed with calculations in the other. Various details of the MQDT model are passed to the module through this initialisation call, as well as a list of experimental energies and experimental errors, indices identifying the $j$ and $k$ channels, and variables controlling either how the theoretical energies are calculated or how the energy dependence of the MQDT parameters is taken into account.

Initialisation must be followed by a call to {\tt search\_intervals}, to set the mesh used in calculations of energies. A list of MQDT parameters must also be passed to the program, i.e., either the $\mu_i^{(0)}$'s, $\mu_i^{(1)}$'s and $\theta_m$'s for eigenchannel calculations, which are passed through {\tt mqdtparams\_eigenchannel}, or the $K_{ij}^{(0)}$'s and $K_{ii}^{(1)}$'s for ${K}$-matrix calculations, which are passed through {\tt mqdtparams\_Kmatrx}.

The module provides are a number of possibilities after this initialisation stage:
\begin{itemize}
    \item call the function {\tt optimizeparams}, which optimises the parameters of the model;
    \item call the function {\tt print\_energies}, which calculates and writes out theoretical energies;
    \item call the function {\tt list\_Zcoeffsandchannelfractions}, which calculates and writes out the mixing coefficients and channel fractions;
    \item call the function {\tt print\_chi2}, which calculates and writes out the $\chi^2$ and reduced $\chi^2$ values characterising the goodness of fit of the model;
    \item call the function {\tt LuFano\_plot}, which draws a Lu-Fano plot; 
    \item call the function {\tt plot\_channelfractions}, which calculates and plots channel fractions.
\end{itemize}
For convenience, the module also contains the following non-computational user-facing functions:
\begin{itemize}
    \item {\tt reset\_calculation\_method}, which can be used for changing the calculation method after the initialization stage;
    \item {\tt print\_channelparams}, which writes out the channel
    parameters passed to the module;
    \item {\tt print\_mqdtparams}, which writes out the MQDT parameters obtained by or passed to the program.
\end{itemize}
More information about these functions can be found in the  user manual joined with this distribution.

The following details should be noted:
\begin{enumerate}
    \item The experimental energies, experimental errors, ionisation limits and calculated energies are generally passed to or from the module in terms of the corresponding wave numbers expressed in $\mbox{cm}^{-1}$.
    \item The channels are labelled in numerical order, starting at 1 as it is customary in the field. Python arrays containing the corresponding information must be dimensioned accordingly. For example, the array {\tt Ilim} containing the ionisation limits of a 2-channel model needs to be created as containing at least three elements, e.g., through the statement {\tt Ilim = numpy.empty(3)}, so that {\tt Ilim[1]} can be set to $I_1$ and {\tt Ilim[2]} to $I_2$ ({\tt Ilim[0]} is not used).
    \item The MQDT parameters are passed between many of the functions of this module as a single 1D array rather than as whole matrices or separate arrays of eigen quantum defects and rotation angles. The way in which these parameters are arranged in this 1D array is not important for routine use of this library.
\end{enumerate}

\section{Examples}
\label{section:examples}
Two example programs are included in this distribution, namely one illustrating the use of the module for calculating energies and producing Lu-Fano plots in the eigenchannel formulation of MQDT, and one illustrating its use for optimizing the parameters of a ${K}$-matrix based model and calculating the resulting energies, mixing coefficients and channel fractions.

The first program (file {\tt ytterbium.py}) concerns the 6snd$\,^1$D$_2$ and 6snd$\,^3$D$_2$ series of $^{174}$Yb. It uses the same model and the same experimental energies and experimental errors as in Tables~V and VIII of Ref.~\cite{Lehec2018} and reproduces the theoretical energies and Lu-Fano plot of Table~V and Fig.~4(a) of that reference. The data are read from the file {\tt ytterbiumD2data.dat}, also included in this distribution. A copy of the output can be found in the files {\tt ytterbium\_output.txt} and {\tt LuFanoplot.pdf}.

The other program (file {\tt strontium.py}) concerns the
5snp$\,^1$P$_1$ series of $^{88}$Sr \cite{Potvliege2023}. A copy of the output is also joined with the distribution, namely the files {\tt strontium\_output.txt} and {\tt channelfractions.pdf}.

\section*{Declaration of competing interest}
The author has no competing financial interests or personal relationships that could have influenced or may appear to have influenced the work reported in this article.

\section*{Acknowledgements}
The author acknowledges very useful discussions with C.~L.~Vaillant and M.~P.~A.~Jones about the application of multichannel quantum defect theory to the calculation of atomic properties, which have formed the basis of the present work. The graphical design of the Lu-Fano plots and plots of channel fractions produced by this software closely follow that of figures presented in Refs.~\cite{Vaillant2014} and \cite{Vaillantthesis}.

\appendix
\section{About the $C_i$ and $A_\alpha$ coefficients}
\label{section:appendix}
The relations between the $C_i$ and $A_\alpha$ coefficients stated in Section~\ref{section:theory} are derived in this appendix. The discussion follows Ref.~\cite{Cooke1985}.

We start with Eqs.~(\ref{eq:Falpha}), (\ref{eq:Galpha}) and (\ref{eq:condition}). Identifying the terms in
$f(\epsilon_i,l_i,r)$ and those in $g(\epsilon_i,l_i,r)$ in this last equation yields, respectively,
\begin{align}
C_i \cos(\pi\nu_i) &= \sum_\alpha U_{i\alpha}\cos(\pi\mu_\alpha)A_\alpha
\label{eq:Ccos}
\end{align}
and
\begin{align}
C_i \sin(\pi\nu_i) &= -\sum_\alpha U_{i\alpha}\sin(\pi\mu_\alpha)A_\alpha.
\label{eq:Csin}
\end{align}
Multiplying Eq.~(\ref{eq:Csin}) by $-i$, adding the result to Eq.~(\ref{eq:Ccos}) and rearranging gives
\begin{equation}
C_i = \sum_\alpha U_{i\alpha}\exp[i\pi(\nu_i + \mu_\alpha)]\,A_\alpha.
\label{eq:Aexp}
\end{equation}
Using the orthogonality of the matrix $[U_{i\alpha}]$, one can also deduce, from this last equation, that
\begin{equation}
A_\alpha = \sum_i U_{i\alpha}\exp[-i\pi(\nu_i + \mu_\alpha)]\,C_i.
\label{eq:Cexp}
\end{equation}
We choose the wave function $\Psi(X,r)$ to be real, which is not restrictive and implies that the coefficients $A_\alpha$ and $C_i$ have a zero imaginary part. As the $U_{i\alpha}$'s are also real, Eqs.~(\ref{eq:CfromA}) and (\ref{eq:Aalphasystem}) then follow from taking the real and imaginary parts of Eq.~(\ref{eq:Aexp}), and  Eqs.~(\ref{eq:AfromC}) and (\ref{eq:Cisystem}) follow from taking the real and imaginary parts of Eq.~(\ref{eq:Cexp}).

Eq.~(\ref{eq:Kmatrixsystem}) follows from re-writing Eq.~(\ref{eq:Cisystem}) in the form
\begin{equation}
    \sum_j U_{j\alpha}\cos(\pi\mu_\alpha)[\tan(\pi\nu_j)+\tan(\pi\mu_\alpha)] \cos(\pi\nu_j)C_j = 0, 
    \quad \alpha=1,\ldots,{N}_{\rm ch},
\end{equation}
dividing each equation by $\cos(\pi\mu\alpha)$, multiplying it on the right by $U_{i\alpha}$, and summing over $\alpha$. Doing so indeed yields
\begin{equation}
\sum_{j,\alpha} U_{j\alpha}
\tan(\pi\nu_j)+\tan(\pi\mu_\alpha)] U_{i\alpha}c_j = 0
\end{equation}
with $c_j = \cos(\pi\nu_j)\,C_j$.
Eq.~(\ref{eq:Kmatrixsystem}) follows since
\begin{equation}
   \sum_{\alpha} U_{j\alpha}\tan(\pi\mu_\alpha)\,U_{i\alpha} = 
    \sum_{\alpha} U_{i\alpha}\tan(\pi\mu_\alpha)\,U_{j\alpha} =
   K_{ij}
\end{equation}
by definition of the $[K_{ij}]$ matrix, and
\begin{equation}
    \sum_{\alpha} U_{j\alpha}U_{i\alpha} = \delta_{ij}
\end{equation}
owing to the orthogonality of $[U_{i\alpha}]$.











\end{document}